\input harvmac  
\lref\soib{ L. L. Vaksman and Ya. S. Soibelman
{\it The Algebra of functions on the quantum group
  $SU(n+1)$ and odd-dimensional quantum spheres,}
 Leningrad Math. J. Vol. {\bf 2} ( 1991 ), No. 5. } 
\lref\ks{ S. Kachru, E. Silverstein, {\it 4d Conformal Field Theories and
Strings on Orbifolds}, Phys.Rev.Lett. {\bf 80} (1998) 4855-4858. }
\lref\law{A. Lawrence, N. Nekrasov, C. Vafa, {\it  On Conformal Theories
in Four Dimensions}, Nucl.Phys. {\bf B533} (1998) 199-209.}
\lref\ozte{Y. Oz, J. Terning, {\it Orbifolds of $AdS_5 \times S^5$ and 4d
Conformal Field Theories}, Nucl.Phys. {\bf B532} (1998) 163-180.}
\lref\gukov{ S. Gukov, {\it Comments on \N=2 AdS Orbifolds},  Phys.Lett.
{\bf B439} (1998) 23-28.}
\lref\ncg{ A. Jevicki, S. Ramgoolam, {\it Non-commutative gravity from
the AdS/CFT correspondence}, JHEP {\bf 9904} (1999) 032.}
\lref\reshtu{ N. Reshetikhin, V. G. Turaev, 
``Invariants of three manifolds  via 
link polynomials and quantum groups'', Invent.Math.103:547-597,1991.}
\lref\frt{ L. D. Faddeev, N. Yu. Reshetikhin, L. A. Takhtadzhyan,
{\it Quantization of Lie groups and Lie algebras}, Leningrad Math.
J. {\bf 1} (1990) 193.}
\lref\malstro{J. Maldacena, A. Strominger, {\it AdS3 Black Holes and a
Stringy Exclusion Principle},  JHEP {\bf 9812} (1998) 005.}
\lref\malda{ J. Maldacena, {\it The Large N Limit of Superconformal Field
Theories and Supergravity}, Adv.Theor.Math.Phys. {\bf 2} (1998) 231-252.}
\lref\rev{ O. Aharony, S. S. Gubser, J. Maldacena, H. Ooguri, Y. Oz, {\it 
Large N Field Theories, String Theory and Gravity},  hep-th/9905111.} 
\lref\jimb{ M. Jimbo, {\it A q-analog of U(gl(N+1) Hecke Algebras and the
Yang-Baxter Equation,}  Lett. Math. Phys. {\bf 11} (1986), 247 - 252.}
%\lref\furl{P. Furlan, L. K. Hadjivanov, I. V. Todorov, {\it Canonical
%Approach to the Quantum WZNW MOdel}, ICTP Trieste preprint IC/95/74, ESI
%234 (1995).}
\lref\furl{P. Furlan, A. Ch. Ganchev, V. B. Petkova, 
             {\it Quantum Groups and Fusion Rule Multiplicities,}
              Nucl. Phys. {\bf B343} (1990) 205.}  
\lref\sal{V. Pasquier, H. Saleur,{\it Common Structures  Between
Finite Systems and Conformal Field through Quantum Groups},
 Nucl.Phys. {\bf B330} (1990) 523.}
\lref\Stein{ H. Steinacker, {\it Unitary Representations of
    Noncompact Quantum Groups at Roots of Unity},
    math.QA/9907021; {\it Unitary Representations and BRST Structure
    of the Quantum Anti-de Sitter Group at Roots of Unity},
    q-alg/9710016; {\it Quantum Groups, Roots of Unity and Particles
    on Quantized Anti-de Sitter Space}, hep-th/9705211;
    {\it Finite Dimensional Unitary Representations of Quantum
    Anti-de Sitter Groups at Roots of Unity}, q-alg/9611009. }
\lref\Keller{ G. Keller, {\it Fusion Rules of $U_qSL(2,C)$, $q^m = 1$,}
                Lett. Math. Phys. {\bf 21}: 273, 1991. } 
\lref\witten{ E. Witten, 
{\it Anti-de-Sitter space and holography,} hepth-9802150. } 
\lref\gkp {S. S. Gubser, I. R. Klebanov and A. M. Polyakov, 
   {\it Gauge Theory Corellators from Non-Critical String Theory},
    Phys.Lett. {\bf B428} (1998) 105-114.}
\lref\salstra{ A. Salam, J. Strathdee, {\it On Kaluza-Klein Theory,}
                   Annals in Phys. {\bf 141}, 316, 1982. } 
\lref\baracz{ A. O. Barut and R. Raczka, 
 { \it Theory of Group Representations and Applications}
PWN, 1980. } 
\lref\charpres{ V. Chari, 
A. Pressley, { \it A guide to quantum groups,} 
CUP, 1994. } 
\lref\bv { M. Berkooz and H. Verlinde, {\it Matrix Theory, AdS/CFT and
Higgs-Coulomb Equivalence}, hep-th/9907100.}
\lref\ho{ C.-S. Chu, P.-M. Ho, H. Steinacker, {\it q-deformed Dirac
Monopole With Arbitrary Charge}, Z.Phys. {\bf C71} (1996) 171-177.}
\lref\majid{T. Brzezinski and S. Majid, {\it Quantum Group Gauge
Theory on Quantum Spaces} Commun.Math.Phys. {\bf 157} (1993) 591. }
\lref\connes{A. Connes, {\it Non-commutative geometry, } 
       Academic Press 1994}              
\lref\Schw{N. Nekrasov, A. Schwarz, {\it Instantons on Noncommutative
$\R^4$ and $(2,0)$ Superconformal Six Dimensional Theory},
hep-th/9802068.}

\def\Rh {\hat{R}}
\def\zb {\bar{z}}
\def\N {{\cal N}}
\def\R {{\bf R}}
\def\C {{\bf C}}

%\draftmode
\nolabels

%\vskip 1cm
 
 \Title{ \vbox{\baselineskip12pt\hbox{  Brown Het-1190 }}}
 {\vbox{
\centerline{ Quantum Spacetimes and Finite N Effects  }
\centerline{ in 4D Super Yang-Mills Theories   }  }}

\centerline{$\quad$ {Pei-Ming Ho$^{1}$,  Sanjaye Ramgoolam$^{2}$
                and Radu Tatar$^{2}$ }}
\smallskip
\centerline{{\sl $^{1}$ National Taiwan University, }}
\centerline{{\sl Taipei 10764, Taiwan, R.O.C. }}
\centerline{{\sl $^{2}$ Brown  University}}
\centerline{{\sl        Providence, RI 02912 }}
\centerline{{\tt pmho@phys.ntu.edu.tw, ramgosk@het.brown.edu,
tatar@het.brown.edu}}
 \vskip .3in

The truncation in the number of
single-trace  chiral primary operators 
of $\N=4$ SYM and its conjectured connection with gravity 
 on quantum spacetimes are  elaborated. 
The model of quantum spacetime we use 
is $AdS^5_q \times S^5_q$ for $q$ a root of unity. The quantum sphere 
 is defined as
a homogeneous space with manifest 
 $SU_q(3)$ symmetry, but as anticipated from the field theory 
correspondence, we show that there is a hidden $SO_q(6)$
symmetry in the constrution. We also study some properties 
 of quantum space quotients as candidate models for 
 the quantum spacetime relevant for some $Z_n$ quiver 
 quotients of the $\N=4$ theory which break SUSY 
 to $\N=2$. We find various qualitative agreements
 between the proposed models and the properties of the 
 corresponding finite $N$ gauge theories.
 
\Date{9/99}
%\draftmode
 
\newsec{ Introduction and summary} 
 
The  Maldacena duality
\refs{ \malda, \gkp, \witten, \rev }  gives a relation 
 between type IIB string theory
on $AdS_{5} \times S^5$ and the $\N = 4$ superconformal four
dimensional super Yang-Mills. 
The gauge group of the field theory is 
$SU(N)$ when the flux through $S^5$ is $N$.
 In the case of
large N and large effective coupling, Maldacena's conjecture relates
the corresponding field theory and the classical supergravity.
Finite $N$ effects contain important information about 
the qualitative novelties of quantum gravity compared to 
classical gravity. 
 
It was proposed in \ncg\ 
that the quantum corrections in the $AdS \times S$ background
 have the effect of deforming spacetime 
 to a non-commutative manifold. The concrete model 
 studied there was $AdS_3 \times S^3$ 
 where the group structure of the manifold 
 allowed a simple non-commutative candidate by
 using quantum groups. An important part  of the  evidence
 was a quantum group interpretation 
 of the cutoff on single particle chiral primaries, 
 first studied under the heading of 
 ``stringy exclusion principle'' in \malstro.  
 
 Here we  develop the same line of argument 
 to understand analogous cutoffs 
 in the spectrum  of chiral primaries 
 of $\N=4$ super Yang-Mills. 
 The cutoffs originate from the 
 fact that the $U(N)$ invariants of the form 
 $tr \Phi^l$, where $\Phi$ is a matrix in the adjoint 
 representation,  are not independent for all 
 values of $l$ and the set of independent invariants 
 truncates at $N$. Since the Yang-Mills theory 
 has an $SO(6)$ global symmetry, we get an $SO(6)$ 
 covariant cutoff on the chiral primaries. 
In the large $N$ limit,  the chiral 
primaries are matched on the gravitational side 
 with modes coming form KK reduction on $S^5$.   
 To understand the cutoff at finite $N$,  we postulate that 
 the $S^5$ is deformed to a quantum sphere in the following way:
 $S^5$ has a 
 description as a coset space $SU(3)/SU(2)$ which generalizes 
 to the q-deformed case as $ SU_q(3)/SU_q(2)$. 
 This construction is shown to have $SO_q(6)$ hidden symmetry 
 by mapping it to another construction 
 of a q-sphere based on \frt,  thus explaining 
 why KK reduction
 on it gives a truncated set of reps. of $SO(6)$. 
 
 We then discuss $\N=2$ theories 
 obtained by taking  $Z_n$ quotients of $U(Nn)$  theories
 with $\N=4$ supersymmetry, which are dual to gravity  
 on $AdS_5 \times S^5/Z_n$. We discuss the cutoffs
 in the spectrum of chiral primaries in the quotient theories. 
 The main point 
 is that, if we ignore states coming from 
 twisted sectors associated with non-trivial 
reps. of $Z_n$, the cutoffs occur at $Nn$ as in the parent 
 theory. To find  candidates for the quantum space dual
 of the quiver theory we identify appropriate  
 automorphisms of the dual theory, which are used to   
 quotient the quantum $S^5_q$ to give a space 
 with $SU_q(2) \times U_q(1)$ symmetry.
 When twisted sectors are taken into account, 
 the cutoff in some chiral primaries charged under the 
$U(1)$  happens not at $Nn$ 
 but at $N$. We  begin a discussion of the quantum space 
explanation of this change of 
 cutoffs. This requires the description 
 of the $S^5_q$ as an $S^1$ fibration over a 
 q-deformed ball, which is acted upon by the 
 quotient.

\newsec{Truncation of generating
  chiral primary operators in $\N = 4$ super Yang-Mills }
 
 Consider the chiral primaries 
 of this theory which are of the form : 
\eqn\chpr{  C_{a_1 a_2 \cdots a_l } tr ( \Phi^{a_1} \cdots \Phi^{a_l} ) } 
 where the $C$ are traceless symmetric tensors
 of $SO(6)$. 
 These symmetric tensors can be decomposed
 under $ SU(3)$ and contain the symmetric rep. 
 of $SU(3)$ corresponding to a Young tableau 
 with one row of length $l$. The polynomials 
 corresponding to this Young tableau are 
 $C_{i_1 i_2 \cdots i_l }
   tr ( \Phi^{i_1} \Phi^{i_2} \cdots \Phi^{i_l} ),  $
where the $C$ are symmetric, and the  $i_k$ are indices running from
 $1$ to $3$ and the corresponding scalars are complex.

It is useful in the discussion of cutoffs to decompose 
 the invariant polynomials in reps. of $SU(3)$. 
In order to obtain the decomposition of symmetric representations of
$SO(6)$ into representations of $SU(3)$, we can  use the
isomorphism between $SO(6)$ and $SU(4)$ under which the vector
representation of $SO(6)$ goes into the antisymmetric representation of 
$SU(4)$, and each symmetric traceless representation of $SO(6)$ goes into
the $(0, k, 0)$ representation of $SU(4)$. The branching rules for the
$(0, k, 0)$ representations of $SU(4)$ into representations of $SU(3)$ 
are, for example, 
\eqn\brules{\eqalign{
& \hbox{ For $k=1$ }\quad  {\bf 6 \rightarrow 3 \oplus \bar{3}},  \cr 
& \hbox{ For $k=2$ } \quad {\bf 20 \rightarrow 6 \oplus
\bar{6} \oplus 8}, \cr 
& \hbox{ For $k=3$ } \quad {\bf 50} \rightarrow { \bf 10 + \bar{10} + 15 +
\bar{15}}. \cr }}
So the vector representation of $SO(6)$
gives a fundamental and an
anti-fundamental representation
corresponding to  $tr (\Phi^i)$ and $tr (\Phi^{* i})$
(i=1,2,3),
which include the
chiral primary operator of dimension 1. 
 The symmetric traceless representation ${\bf 20}$ gives the following
 operators : $tr (\Phi^{( i_{1}} \Phi^{i_{2})})$
(the ${\bf 6}$ representation), $tr (\Phi^{* ( i_{1}} \Phi^{* i_{2})})$  
(the ${\bf \bar{6}}$ representation), $tr (\Phi^{( i_{1}} \Phi^{*
i_{2})})$
(the ${\bf 8}$
representation). These include the  chiral primary
operator of dimension 2. The representation ${\bf 50}$ of $SO(6)$ gives
the following : $tr (\Phi^{( i_{1}} \Phi^{i_{2}}
\Phi^{i_{3})})$ (the ${\bf 10}$ representation), its conjugate one
(involving the complex conjugate fields and corresponding to the
${\bf \bar{10}}$ representation),  $tr (\Phi^{( i_{1}} \Phi^{* i_{2}} 
\Phi^{* i_{3})})$ (the ${\bf 15}$ representation) and its complex conjugate
(the ${\bf \bar{15}}$ representation). These  contain  the chiral primary
of dimension 3. We can continue the discussion to show that 
each symmetric traceless rep. of $SO(6)$ contains a chiral 
primary belonging to a rep. of $SU(3)$ associated to a symmetric
Young tableau.

 The chiral primary
 operators are not all independent at  
finite $N$. Consider first the operators which look like 
$tr (\Phi^{N+1})$ when the gauge group is $U(N)$. If the generators of the
Lie algebra are $T^{a}$, the chiral primary operator is written as 
$tr (\Phi_{a} T^{a})^{N+1} = \Phi_{a_{1}} \cdots \Phi_{a_{N+1}}
tr (T^{a_{1}} \cdots T^{a_{N+1}})$. But $tr (T^{a_{1}} \cdots
T^{a_{N+1}})$ 
is just the $C_{N+1}$ Casimir operator of $U(N)$ and we know that $U(N)$
has
only N independent Casimir operators so $C_{N+1}$ is not independent and
can be written in terms of lower Casimir operators. Therefore, the
conclusion is that $tr (\Phi^{N+1})$ can be written in terms of $tr
(\Phi^{N})$, $tr (\Phi^{N-1})$ and so on. Thus $tr (\Phi^{N+1})$
does not describe a single particle state. The
conclusion is that for the group $U(N)$ we have a truncation on the chiral
primary operators such that the highest power is N.
For instance, $\Phi$ can be one of the three $\Phi_i$'s.
The rest of the symmetric polynomials 
 like $ tr ( \Phi_1^i \Phi_2^j \Phi_3^k )$ can also 
 be decomposed since they can be obtained from the 
 chiral primary by action of $SO(6)$. In fact for any given dimension, 
 there is only one short representation, so that operators of the form 
\eqn\supst{ 
 tr ( F \Phi \Phi \cdots ) } 
can also be shown to be decomposable as they are obtained 
 from the chiral primary by action of the 
SUSY operators. 
 
The result is that we have a truncation on the short representations
for the gauge group $U(N)$, the maximal symmetric traceless representation
of $SO(6)$ being the one with N boxes in the Young tableau.   
This will allow us to identify 
 a candidate quantum sphere relevant for the 
 spacetime understanding of finite $N$ effects. 
 After describing some preliminaries on
 quantum groups we will describe the 
relevant quantum space in section 3.3

\newsec{ Non-commutative spacetime } 
 
Natural non-commutative candidates 
 for  $AdS_5 \times S^5$ are obtained by 
 deforming the coset struture of the spaces 
 involved  using quantum groups. 
 We will be concerned with some 
 detailed properties of the q-deformed $S^5$
 in this paper, which are relevant for the 
 truncation in KK modes.  
 The unit five sphere is the space 
$\sum_{i=1}^{6} x_{i}^2 = 1$ where $x_i$ are coordinates in $\R^6$.
$SO(6)$ acts transitively on the solutions of this equation. 
A point, say $(0,0,0,0,0,1)$ is left fixed
 by $SO(5)$. This allows an identification 
of the  sphere with the  coset $SO(6)/SO(5)$.
It is posible to consider $\R^6$ as a complex space $\C^3$ with coordinates
$z_0 = x_1 + i x_2, z_1 = x_3 + i x_4, z_2 = x_5 + i x_6$ and the sphere
becomes a surface
$\sum_{j=0}^{2} |z_{j}|^2 = 1$ in $\C^3$. In this case the
sphere is seen as a coset $SU(3)/SU(2)$. 
The latter coset space struture 
allows a simple quantum group generalization.

\subsec{ Preliminaries of quantum groups }
 
The standard $q$-deformation of quantum groups is given
in \frt\ . For a matrix ${T^i}_j$ of a quantum group,
the commutation relations among the matrix elements
are given by
\eqn\rtt{ \Rh_{12}T_1 T_2=T_1 T_2\Rh_{12}. }
This is a shorthand of the following
\eqn\nrtt{ \Rh^{ij}_{kl}{T^k}_m{T^l}_n={T^i}_k{T^j}_l\Rh^{kl}_{mn}. }
The matrix elements ${T^i}_j$
in the fundamental rep.
generate the algebra of functions
on the quantum group.
The matrix $T$ has an inverse $T^{-1}$ given by the antipode.
The antipode is an automorphism $S$ of this algebra
such that $S({T^i}_j)=(T^{-1})^i_j$.
 
For $SL_q(N;{\bf C})$, the $\Rh$ matrix is given by
\eqn\hatR{ \Rh^{ij}_{kl}=\delta^i_l\delta^j_k(1+(q-1)\delta^{ij})
+\lambda\delta^i_k\delta^j_l\theta(j-i), }
where $\theta(j-i)=1$ if $j>i$ and $\theta(j-i)=0$ otherwise.
In \frt\ a $\ast$-anti-involution is given for $SL_q(N;{\bf C})$.
With respect to this $\ast$-anti-involution, a real form of
$SL_q(N;{\bf C})$ can be defined by $({T^i}_j)^*=(T^{-1})^j_i$.
For $q$ being a phase, the real form $SL_q(N)$ can be defined
by $T^*=T$; and for $q$ being real, the real form $SU_q(N)$
can be defined.
(We require that the $\ast$ of a complex number be its
complex conjugation, so e.g. $q^*=q^{-1}$ if $q$ is a phase.)
In this paper what we need is $SU_q(N)$,
but there is no $\ast$-anti-involution for this purpose
when $q$ is a phase. It turns out that the appropriate
$\ast$-structure is an involution instead of anti-involution.
(An involution does not reverse the ordering of
a product and an anti-involution does.)
Let
\eqn\g{ {g^i}_j=q^i\delta^i_j, }
where $i,j=0,1,\cdots,N-1$,
then we define
\eqn\star{ T^{\dagger}=g^{-1}T^{-1}g, }
where $T^{\dagger}$ is the transpose of $T^*$.
One can check that this definition gives a $\ast$-involution.
First, because $S(S(T))=g^2 T g^{-2}$, $(T^*)^*=T$.
Secondly, \rtt\
is invariant under the action of $\ast$.
To check this, one can use the following identities
\eqn\Rtrans{ \Rh^{ij}_{kl}=\Rh^{kl}_{ij}, }
\eqn\Rstar{ \Rh(q^{-1})_{12}=\Rh(q)^{-1}_{21}, }

\eqn\Rgg{ g_1^{-1}g_2^{-1}\Rh_{12}g_1 g_2=\Rh_{12}. }
 
\subsec{ The quantum sphere } 
 
To begin we define  the quantum complex plane ${\bf C}_q^N$
which has the symmetry group $SU_q(N)$ acting on it.
The algebra of functions on ${\bf C}_q^N$ is generated
by the coordinates $z_i$, $i=0,1,\cdots,N-1$,
which satisfy the following commutation relations
\eqn\zz{ z_1 z_2=q^{-1}z_1 z_2 \Rh_{12}. }
More explicitly, it is
\eqn\nzz{ z_i z_j=q^{-1}z_k z_l\Rh^{kl}_{ij}. }
The coordinates $z$ transform under an $SU_q(N)$ matrix $T$ as
\eqn\zT{ z\rightarrow z T. }
We let all $z_k$'s to commute with all ${T^i}_j$'s.
Due to \rtt\ ,
the relations \zz\ is preserved by this transformation.
(Note that ${T^i}_j$ commutes with $z_k$ for all $i,j,k$.)
The complex conjugation of $z$ is defined as a $\ast$-involution.
Let $\zb^i=z_i^*$. The $\ast$ of \zz\ is
\eqn\zbzb{ \zb_1\zb_2=q^{-1}\Rh_{21}\zb_1\zb_2, }
which is covariant under the transformation
$\zb\rightarrow T^{\dagger}\zb$, as the $\ast$ of \zT\ .
To complete the definition of the algebra on ${\bf C}_q^N$,
we also need to define how $z$ commutes with $\zb$.
Let
\eqn\zzb{ \zb_1 z_1=q^{-1}z_2 g_2\Rh_{12}g_1^{-1}\zb_2, }
which means
\eqn\nzzb{ \zb^i z_k=q^{-1+j-k}z_j\Rh^{ij}_{kl}\zb^l. }
It is covariant under the action of $SU_q(N)$.
 
The coefficient of $q^{-1}$ on the right hand side of \zzb\
is chosen such that the radius squared
\eqn\rr{ r^2=zg\zb }
is a central element in the algebra.
One can also check that $r^2$ is real: $(r^2)^*=r^2$.
Since $r^2$ commutes with everything else, we can define
a new algebra by the algebra of $z,\zb$, modulo the condition
$r^2=1$. This is the algebra of functions on the quantum
sphere $S_q^{2N-1}$. It can be identified with the quantum sphere
defined as $SU_q(N)/SU_q(N-1)$ \soib\ .
Explicitly, the commutation relations of $z,\zb$ are
\eqn\zizj{ z_i z_j=q z_j z_i, \quad i<j, }
\eqn\zizbj{ z_i\zb^j=q\zb^j z_i, \quad i\neq j, }
\eqn\zbizbj{ \zb^i\zb^j=q^{-1}\zb^j\zb^i, \quad i<j, }
\eqn\zizbi{ z_i\zb^i=\zb^i z_i-q^{-1}\lambda\sum_{j>i}q^{j-i}z_j\zb^j. }
 
Another natural candidate for the q-deformed sphere is
the $SO_q(N)$-covariant quantum Euclidean space ${\bf R}_q^N$
modulo the unit radius condition.
The quantum group $SO_q(N;{\bf C})$ is defined by \rtt\ 
with a different $\Rh$ matrix, which also has the
properties \Rtrans\ and \Rstar.
In addition, one has
\eqn\RCC{ C_1 C_2\Rh_{12}C_1 C_2=\Rh_{21}, }
where ${C^i}_j=\delta^i_{N+1-j}$.
 
For $q$ being a phase, the real form $SO_q(n,n)$ or $SO_q(n,n+1)$
can be defined by $T^*=T$ for a $\ast$-anti-involution.
For real $q$, the real form $SO_q(N;{\bf R})$ exists with respect to
the $\ast$-anti-involution $T^*=GTG^{-1}$,
where ${G^i}_j=q^i\delta^i_{N+1-j}$.
We need $SO_q(N;{\bf R})$ for $q$ being a phase,
so again we can only define it with respect to a $\ast$-involution.
We find the appropriate $\ast$-involution to be given by
\eqn\CTC{ T^*=CTC. }
Incidentally, if one wants to define $SO_q(2n,2m)$,
we can generalize the above to $T^*=\hat{C}T\hat{C}^{-1}$,
where ${\hat{C}^i}_j=\epsilon_i\delta^i_{N+1-j}$ ($N=2(n+m)$)
with $m$ of the $\epsilon$'s equal to $-1$ and the rest to $1$.
This includes $SO(4,2)$ which is of interest in defining the
 deformation of the AdS part. 
The $\ast$-involution in the corresponding universal enveloping
algebra is recently given in \Stein, where some useful information
about unitary representations when $q$ is a root of unity is also given.
In the following we will concentrate on $SO_q(N;{\bf R})$,
which will be denoted as simply $SO_q(N)$.
 
The $SO_q(N)$-covariant algebra of functions on
the quantum Euclidean space is defined by \frt\
\eqn\xx{ x_1 x_2=q^{-1} x_1 x_2\Rh_{12}+
\kappa R^2 G, }
where $\kappa=(1-q^{-2})/(1+q^{N-2})$ and $R^2=x^t Gx$ is the radius
squared.
The transformation of $SO_q(N)$ on $x$ is $x\rightarrow xT$.
The $\ast$-involution compatible with \CTC\ is
\eqn\xstar{ x^*=xC. }
 
Since we have demanded all algebras to have the $\ast$-involution,
it follows that there is a symmetry of $q\rightarrow q^{-1}$
if we simultaneously take $x\rightarrow x^*$, $T\rightarrow T^*$ etc.
It is therefore equivalent to say that we have $S_q^5$ or $S_{q^{-1}}^5$.
 
It can be explicitly checked that the algebra of ${\bf C}_q^3$
is the same as the algebra of ${\bf R}_q^6$ via the identification
$z_i=x_{i+1}$ and $\zb^i=x_{6-i}$ for $i=0,1,2$,
and then $R^2=(q^2+q^{-2})r^2$.
Therefore the three definitions of $S_q^5$ are actually equivalent:
$SU_q(3)/SU_q(2)={\bf C}_q^3/(r^2=1)={\bf R}_q^6/(R^2=q^2+q^{-2})$.
In the first two models of $S_q^5$, the action of $SU_q(3)$
is manifest. In the third model the action of $SO_q(6)$ is manifest.
This suggests that $SU_q(3)$ can be realized as a subgroup of
$SO_q(6)$. This is indeed the case.
Given a $3\times 3$ $SU_q(3)$ matrix $t$,
we can define a special $6\times 6$ $SO_q(6)$ matrix $T$ by
$T_{ij}=t_{ij}$ for $i,j=1,2,3$,
$T_{ij}=t^*_{(7-i)(7-j)}$ for $i,j=4,5,6$,
and all other elements $T_{ij}=0$.
 
\subsec{ Quantum sphere for $U(N)$ $\N=4$ SYM } 
 Now that we have established 
the $SO_q(6)$
symmetry of the quantum sphere, we use the fact that KK 
 reduction on this space will give 
 a family of reps. of $SO_q(6)$.
In Sec. 3.5 we will show that $SO_q(6)$ can be identified with
$SU_q(4)$.
 For $q$ being a root of unity, the reps. of 
 $SU_q(4)$  contain 
 indecomposable reps. which form an ideal under tensor 
 products. After quotienting these out, 
 one is left with a set of standard 
 reps. with a truncation. 
 The length of the first 
 row of the Young tableau cannot exceed
$k$ for $ q = e^{i \pi \over { k+4 }}$. This fact 
 is familiar from 2D WZW models \refs{ \sal,  \furl }
 and has been studied in detail for $U_qSU(2)$ 
 in \Keller\reshtu.   
 This allows us to identify $k \sim N $ 
 to get a quantum sphere which gives a Kaluza-Klein reduction
 which agrees with the spectrum of chiral primaries discussed in section
 2.

\subsec{ $Z_n$ automorphisms and symetries of quantum 
          quotient spaces }

 We now discuss some sub-algebras
 relevant for the quantum space 
 analog of the $\N=2 $ quotient theories which will 
 be discussed in the next section. 
The quantum group $SU_q(3)$ has $SU_q(2)$ as a subgroup.
Given a $2\times 2$ $SU_q(2)$ matrix $t$,
we can define a $3\times 3$ $SU_q(3)$ matrix $T$ by
$T_{ij}=t_{ij}$ for $i,j=1,2$, $T_{33}=1$
and all other elements $T_{ij}=0$.
The group $Z_n$ mentioned in previous sections
can then be embedded in $SU_q(2)$
as a diagonal matrix diag$(\omega,\omega^{-1})$, where $\omega^n=1$.

Classically, $SO(6)$ has a maximal subgroup 
$SU(2)\times SU(2)\times U(1)$. This is also true for $SO_q(6)$. 
 The presence of a $U_q (SU(2) ) \times U_q (SU(2) ) \times U_q ( U(1) )$
  subalgebra in $U_q ( SO(6)) $ or $U_q ( SU(4) ) $ is clear from the 
 definition of these algebras using 
 the q-analog of the Chevalley-Serre basis \jimb. 
 From the point of view of the algbera of functions 
 on $SO_q(6)$ we can describe a $SU_q(2) \times SU_q(2) \times U_q(1)$ 
 as follows. 
 The first $SU_q(2) $ is 
\eqn\frssu{ 
 T_{mn} = \pmatrix { & \alpha & \beta & 0 & 0 & 0 & 0 \cr 
           & \gamma & \delta & 0 & 0 & 0 & 0 \cr 
           & 0 & 0 & 1 &  0 & 0 & 0 \cr
           & 0 & 0 & 0 &1 &  0 & 0 \cr 
           & 0 & 0 & 0 & 0 & \alpha & - \beta \cr 
           &  0 & 0 & 0 & 0 & -\gamma & \delta \cr }. } 
The second $SU_q(2) $ is 
\eqn\secs{ 
T_{mn} = \pmatrix { & a  & 0 & 0 & 0 &  b & 0 \cr 
           & 0 & a & 0 & 0 & 0 & -b \cr 
           & 0 & 0 & 1 &  0 & 0 & 0 \cr
           & 0 & 0 & 0 &1 &  0 & 0 \cr 
           & c & 0 & 0 & 0 & d & 0 \cr 
           &  0 & -c & 0 & 0 &0 & d \cr }. }
The $U_q(1)$ is 
\eqn\tu{ 
T_{mn} = \pmatrix { & 1 & 0 & 0 & 0 &0  & 0 \cr 
           & 0 & 1 & 0 & 0 & 0 & 0 \cr 
           & 0 & 0 & \Lambda &  0 & 0 & 0 \cr
           & 0 & 0 & 0 & \Lambda^{-1}  &  0 & 0 \cr 
           & 0 & 0 & 0 & 0 & 1 & 0 \cr 
           &  0 & 0 & 0 & 0 &0 & 1 \cr }. }
 These $6\times 6$ matrices corresponding
to the three subgroups $SU_q(2)$, $U_q(1)$ and $SU_q(2)$
commute with one another, and
they all satisfy
 the relation \rtt\ for $SO_q(6)$.
Note that in checking whether these matrices
commute with one another, we should take
any entry in a matrix to commute with any entry
in another matrix.
The reason for this is that if we take
two matrices $T$ and $T'$ from a quantum group,
for their product to satify the same RTT relation \rtt\
we should let all entries in $T$ to commute with
all entries in $T'$.
(The functions on two copies of the group manifold
is given in a tensor product.)
 
The $Z_n$ symmetry we quotient by is 
 a subalgebra of 
 $U_q(1)$ which is embedded as
$T_{ij}=T_{(i+4)(j+4)}=$
diag$(\omega,\omega^{-1})$ for $i,j=1,2$,
$T_{33}=T_{44}=1$,
and all other elements vanishing,
for $|\omega|=1$.
% The other $U(1)$ is given by
% $T_{33}=\Lambda$, $T_{44}=\Lambda^{-1}$,
% $T_{ii}=1$ for $i=1,2,5,6$,
% and all other elements vanishing,
% for $|\Lambda|=1$.
% The factor of $SU_q(2)$ is embedded in $SO_q(6)$ by
% letting $T_{11}=T_{22}=a$, $T_{15}=-T_{26}=b$,
% $T_{51}=-T_{62}=c$, $T_{55}=T_{66}=d$,
% $T_{33}=T_{44}=1$ and all other elements vanishing,
% for an $SU_q(2)$ matrix with elements $(a,b,c,d)$.
 
The commutant of the $Z_n$ action 
 is $SU_q(2) \times U_q(1) $, with the same 
 $q$ as before. This fact will be useful in 
 a quantum space-time understanding of the 
relation between cutoffs on chiral primaries 
 in an $\N=4$ $U(Nn)$ theory and its $Z_n$ quotient.

% This predicts a cutoff at 
% $ \sim Nn$ for the quotient theories. 
% This agrees with the cutoffs in the 
% field theory, to be discussed in the next section, 
% if we ignore twisted sectors. A more elaborate 
% construction should be invoked for the twisted 
% sectors, which we will come back to in the final section. 

\subsec{  $SU_q(4)$  symmetry }

Since the universal enveloping algebra $U_q(G)$
for a classical group $G$ is completely determined
by the Cartan matrix of $G$, and since the Cartan 
Matrix of $SU(4)$ is the same as that for
 $SO(6)$, 
$SO_q(6)$ is identical to $SU_q(4)$ ( up to 
global differences which do not affect the general sub-group 
 structure). 
Therefore $SO_q(6)$ has a subgroup $SU_q(3)$
which is manifest in the $SU_q(4) $ description. 
%up to a redefinition of $q$.
%Since both $SO_q(6)$ and $SU_q(4)$ have
%the subgroup $SU_q(3)$, we simply have
%$SO_q(6)\simeq SU_q(4)$.

  While we arrived at the  $SU_q(4) \sim SO_q(6)$ 
 symmetry  by explicitly mapping to an algebra which had the larger
 symmetry, we can also guess it by an indirect argument. 
 A hint  for the hidden $SU(4)_q$ comes 
 from a consideration of  KK reduction on 
 $SU_q(3)/SU_q(2)$. Suppose we are dimensionally 
 reducing a scalar on the coset $SU(3)/SU(2)$. We have to look for all 
 reps. of $SU(3)$ which contain a scalar of $SU(2)$ 
 \salstra. We know that the reps. of $SU(3)$ we have
 are precisely such that they combine into reps. of $SU(4)$. 
 To KK reduce on $ SU_q(3)/SU_q(2)$, we q-deform 
 the rule above and look for reps. of $SU_q(3)$ which 
 contain the scalar of $SU_q(2)$. It is very plausible 
 that the reps. still combine into reps. of $SU_q(4)$
 since the structure of the reps. at roots of unity  
 remains the same as long as we stay within the cutoff. 
 This can be proved by using  
 the generalization of the Gelfand-Zetlin bases
 ( described for example in \baracz), 
 which exists because of some special properties  
 of the branching rules in the sequence of subgroups
%h:
 $ SU(2) \subset SU(3) \subset SU(4)$. The Gelfand-Zetlin bases 
 have been generalized to roots of unity ( discussed for example
 in \charpres\ ) 
 so this should provide the proof that the desired 
q-generalization of the branching rules is correct.   

The matching between $SU_q(4)$ and $SO_q(6)$ can
also be illustrated as follows.
There is an $SU_q(3)$ subgroup of $SU_q(4)$ which
acts on $(z_1, z_2, z_3)$. Let us represent it as
\eqn\su{
\pmatrix{
& a & b & c & 0 \cr
& d & e & f & 0 \cr
& g & h & p & 0 \cr
& 0 & 0 & 0 & 1 \cr}.}
This corresponds to the $SU_q(3)$ embedded in $SO_q(6)$
which is described in a previous section.
 
In addition to the $SU_q(2)$ subgroups in this $SU_q(3)$,
there are three other ways of embedding $SU_q(2)$ in $SU_q(4)$:
\eqn\susu{
\pmatrix{
& a & 0 & 0 & b \cr
& 0 & 1 & 0 & 0 \cr
& 0 & 0 & 1 & 0 \cr
& c & 0 & 0 & d \cr}, \quad
\pmatrix{
& 1 & 0 & 0 & 0 \cr
& 0 & a & 0 & b \cr
& 0 & 0 & 1 & 0 \cr
& 0 & c & 0 & d \cr}, \quad
\pmatrix{
& 1 & 0 & 0 & 0 \cr
& 0 & 1 & 0 & 0 \cr
& 0 & 0 & a & b \cr
& 0 & 0 & c & d \cr},}
where $a,b,c,d$ are the four functions which constitute
an $SU_q(2)$ matrix in the fundamental representation.
These three $SU_q(2)$ subgroups all commute with one another. They
are characterized by their being commuting with
different $SU_q(2)$ subgroups of the $SU_q(3)$ mentioned above.
 
It is now not difficult to guess what their correspondence
in $SO_q(6)$ is.
The corresponding $SU_q(2)$ matrices are:
\eqn\suso{
\pmatrix{
& 1 & 0 & 0 & 0 & 0 & 0 \cr 
& 0 & a & 0 & b & 0 & 0 \cr 
& 0 & 0 & a & 0 &-b & 0 \cr
& 0 & c & 0 & d & 0 & 0 \cr 
& 0 & 0 &-c & 0 & d & 0 \cr 
& 0 & 0 & 0 & 0 & 0 & 1 \cr},
\pmatrix{
& a & 0 & 0 & b & 0 & 0 \cr 
& 0 & 1 & 0 & 0 & 0 & 0 \cr 
& 0 & 0 & a & 0 & 0 &-b \cr
& c & 0 & 0 & d & 0 & 0 \cr 
& 0 & 0 & 0 & 0 & 1 & 0 \cr 
& 0 & 0 &-c & 0 & 0 & d \cr},}

and

\eqn\susoi{
\pmatrix{
& a & 0 & 0 & 0 & b & 0 \cr 
& 0 & a & 0 & 0 & 0 &-b \cr 
& 0 & 0 & 1 & 0 & 0 & 0 \cr
& 0 & 0 & 0 & 1 & 0 & 0 \cr 
& c & 0 & 0 & 0 & d & 0 \cr 
& 0 &-c & 0 & 0 & 0 & d \cr}. }
 
The matching above takes care of the 8 generators in
the $SU_q(3)$ and 2 generators in each of the three $SU_q(2)$.
Since there are a total of 15 generators in $SU_q(4)$ or $SO_q(6)$,
there is still one generator left,
the $U(1)$ generator which can be represented as
\eqn\usu{
\pmatrix{
& \Lambda & 0 & 0 & 0 \cr 
& 0 & \Lambda & 0 & 0 \cr 
& 0 & 0 & \Lambda & 0 \cr
& 0 & 0 & 0 & \Lambda^{-3} \cr}, \quad
\pmatrix{
& \Lambda {\bf 1}_{3\times 3} & 0 \cr
& 0 & \Lambda^{-1} {\bf 1}_{3\times 3} \cr} }
in $SU_q(4)$ and $SO_q(6)$ respectively.
 
\newsec{Chiral primary operators for $\N = 2$ quotient theories}
 
\subsec{Conformal field theory discussion } 
Maldacena's conjecture has been extended to the case of orbifolds.
In order to preserve the conformal symmetry, we need to keep the AdS part
untouched and to act with orbifold groups only on $S^5$ 
\refs{ \ks, \law, \ozte , \gukov }.

An $\N = 2$ theory is obtained if  we act with a $Z_n$ group on 
two out of three complex fields, one of them being left unchanged.
The $Z_n$ quotienting is accompanied by a 
gauge transformation. 
\eqn\quot{\eqalign{ 
&   \Omega \Phi_1 \Omega^{-1} = \omega \Phi_1, \cr  
&   \Omega \Phi_2 \Omega^{-1} = \omega^{-1} \Phi_2, \cr
&   \Omega \Phi_3 \Omega^{-1} =  \Phi_3, \cr 
&   \Omega D_A \Omega^{-1}    = D_A, \cr }}
where $D_A$ is the covariant derivative.

The $\Phi$'s are $Nn \times Nn$  matrices. 
 $\Omega$ can be chosen to be
$ diag ( 1, \omega^{-1}, \omega^{-2} 
\cdots \omega^{-(n-1)} )$. 
After taking the quotient the gauge 
 group becomes $SU(N)^{\otimes n}$, with the surviving 
 gauge fields being diagonal $N \times N $ blocks,
 and the bosonic matter content is :  
\eqn\surv{\eqalign{ 
 & \Phi_1 = \pmatrix{ & 0 & Q_1^{(1)}&  0 & 0 & 0 & \cdots  \cr 
                      & 0 & 0 & Q_1^{(2)}&  0&  0& \cdots \cr  
                      & 0 &  0 &0&  Q_1^{(3)} &  0 & \cdots \cr 
                      & \vdots & \vdots & \vdots & \vdots &
                        \vdots & \ddots & \cr }, \cr 
  & \Phi_2 =  \pmatrix{ & 0 & 0&  0 & 0 & \cdots  \cr 
                      & Q_2^{(1)} & 0 &0& 0 & \cdots \cr  
                      & 0 &   Q_2^{(2)} &0& 0 & \cdots \cr 
                      & 0 & 0 &   Q_2^{(3)}& 0 & \cdots \cr
                      & \vdots & \vdots & \vdots & \vdots &
                        \ddots & \cr }, \cr 
 & \Phi_3 = \pmatrix{ & \hat \Phi_3^{(1)} &  0 & 0 & 0 & \cdots  \cr 
                      & 0 &  \hat \Phi_3^{(2)} & 0 & 0 & \cdots \cr  
                      & 0 &  0 & \hat \Phi_3^{(3)} &  0 & \cdots \cr 
                      & \vdots & \vdots & \vdots & \vdots &
                        \ddots & \cr }, \cr }}
where $Q_1^{(i)} $ are fields in the $(N_i, \bar N_{i+1} )$ 
 representation, and $Q_{2}^{(i)} $ are in the 
 $( \bar N_{i}, N_{i+1})$ representation.  
The surviving global symmetry is $SU(2)_R \times U(1)_R \times Z_n$. 
 The pair $( Q_{1}^{(i)}, ( Q_{2}^{(i)})^* )$ 
 is a doublet of $SU(2)_R$ and uncharged 
 under $U(1)_R$, while $ \Phi_3^{(i)}$ are singlets under 
 this $SU(2)_R$ but have charge $1$ under the 
 $U(1)$. The $SU(2)_R$ has a $U(1)$ subgroup, under which the 
 $Q_{1}^{(i)}$ and $Q_2^{(i)}$ have charge $1$. 
 The $Z_n$ acts as  cyclic permutations on the 
 $n$ factors of $SU(N)$ and on the $i$ index of 
 $Q_{1}^{(i)}, Q_{2}^{(i)}, \Phi_{3}^{i}$.  
 Geometrically, the symmetry $SU(2) \times U(1) \times Z_n$
 can be understood, by 
 describing  $S^5/Z_{n}$ as a fibration 
of  $S^2 \times S^1$  over $S^2/Z_n$. 
 
 Out of the chiral primaries of the 
 $\N=4$ theory, those giving non-trivial 
 operators are of the form 
 $tr \Phi_3^{l} $, $tr ( \Phi_{1} \Phi_{2})^l$
 and $tr( \Phi_{1} \Phi_2 \Phi_3^{m} )^{l}$. 
 As in the discussion of cutoffs for 
 the $U(Nn)$ theory we can write traces 
 of powers of these fields in terms of products 
 of traces when the power exceeds $Nn$. 
 For example this leads to 
\eqn\trexp{ tr(\Phi_3 )^{Nn +1 } = tr ( \Phi_3 )^{Nn} tr ( \Phi_3 ) +
\cdots, } 
 where the $\cdots$ stands for other terms 
 involving other splittings of the $(Nn + 1)$'th power. 
 We can rewrite this as follows
\eqn\trexpi{  tr(\Phi_3 )^{Nn +1 } = \sum_{i} tr ( \hat \Phi_3^{(i)}
)^{Nn} 
 \sum_{j} tr ( \hat \Phi_3^{(j)} ) + \cdots. } 
 Note that the splitting involves factors which are separately
 $Z_N$ invariant. 
 The same holds for other operators, 
 e.g. $tr ( \Phi_1 \Phi_{2} )^{Nn+1} $.

 Another kind of splitting  occurs if we allow the factors to 
 come from twisted sectors. This splitting will happen 
 at a different value of the powers, i.e at $N+1$ rather 
 than $Nn+1$. It uses
the
fact that 
\eqn\trexpii{ tr(\hat \Phi_3^{(i)} )^{N +1 } = 
tr(  \hat \Phi_3^{(i)} )^N tr ( \hat \Phi_3^{(i)} )  + \cdots, } 
 and it leads to
\eqn\txp{ tr( \Phi_3 )^{N +1 } = \sum_{i=1}^{n}
 tr(  \hat \Phi_3^{(i)} )^N  tr ( \hat \Phi_3^{(i)} ) + \cdots. } 
Now the factors are not separately invariant. 
 They come from twisted sectors. 
Similar equations can be written for the other operators, 
 for example,
 \eqn\txp{ tr( \Phi_1 \Phi_2 )^{N +1 } = \sum_{i=1}^{n}
 tr(  \hat \Phi_1^{(i)} \hat \Phi_2^{(i)} )^N  
  tr ( \hat \Phi_1^{(i)} \hat \Phi_2^{(i)} ) + \cdots } 
However as discussed in \gukov\ the
factors appearing on the right hand side in this expression
 are not chiral primaries because they appear as derivatives of a
superpotential.

\subsec{ Quantum space explanation of the cutoffs} 

 We saw in the previous subsection
 that if we start with a gauge theory 
 with gauge group $U(Nn)$ and take the $Z_n$ quotient,
 we get a gauge theory with a product of $U(N)$ gauge 
 groups. Restricting attention
 to $Z_n$ invariant operators, i.e the untwisted sector, 
 we find that the cutoff stays at $Nn$. 
 We expect on the gravity side 
 that a discussion ignoring the twisted sectors
 can be simply reproduced by studying a 
 quotient of the q-sphere. 
 We found indeed in the previous section 
 that after quotienting $S^5_q$ by a 
 $Z_n$ element inside the 
 $SO_q(6)$, we are left with a surviving $ SU_q(2) \times U_q(1)$ 
 with the same value of $q$ that we started with. 
 This gives an explanation of the fact that 
 the cutoff $ \sim Nn$ stays at  $ \sim Nn$.

 \subsec{ Comments on twisted sectors } 

 The quantum space explanation 
 of the behaviour of the cutoffs 
 when we include the twisted sectors
 is much more intricate. This is only a 
 preliminary discussion.

 The twisted sector  has to do with 
 dimensional reduction 
 on the singular cycles of $D^4/Z_n$ followed
 by the reduction on $S^1$ \gukov. 
 
To understand the twisted sector
 chiral primaries from the gravity point 
 of view in the large $N$ limit
 one starts with a description of $S^5/Z_n$ as 
 $S^1$ fibred over $B_4/Z_n$ wher $B_4$ is the 4-ball.
 Ten-dimensional gravity 
 is reduced on the $B_4/Z_n$
 and then the KK 
 reduction on the $S^1$ is performed. The cohomology 
 of the blown-up $B_4/Z_n$ space is used to 
determine the type of particles we get \gukov.

 We can try to generalize this discussion 
 to the case of $S^5_q/Z_n$. The twisted sector 
 states are localized in the $Z_1,Z_2$
 direction when the $Z_n$ quotient acts in these 
 directions. The wavefunctions of the twisted sector  
 states are arbitary functions 
 of the phase of $Z_3$  and localized at $ Z_1 = Z_2 = 0 $
 in the case $q=1$. A very intriguing 
 property of the $S^5_q$ we described is that  
 $Z_1=Z_2= 0 $ is not
 compatible with the algebra \zizj\ - \zizbi\ .
 In a sense the origin of the ball is smoothed out. 
 It would be very interesting if this result of
 noncommutativity provides an effective way to
 describe the resolution of the fixed point
 in a similar way as \Schw\ , where it was shown
 that the instanton moduli space can be resolved
 by deforming the base space into a quantum space.
 Another possibility is that
 we could also have chosen to work with a $Z_n$ action 
 on $Z_2,Z_3$. In that case there is a  circle over
 $Z_2=Z_3=0$ on $S_q^5$. So to describe the twisted 
 sector states, we need to extend the algebra 
 $ S^5_q/Z_n$ by adding certain delta functions mutliplied with 
 arbitrary powers of $Z_3$ ( with unit norm ). 
 The results from the field theory suggest 
 that this algebra should admit 
 a consistent truncation which restricts the 
 $U(1)$ charge at order $N$, since we saw 
 that with the cutoffs on single trace operators 
 of the form $tr \Phi_3^l$ happens at order $N$.  
 It will be very interesting to see if the quantum 
 space techniques can reproduce this field theory 
 result. Some relevant techniques on quantum principal bundles may be
found in \refs{\ho,\majid}  and on K-theory of quantum spaces in \connes.

\newsec{\bf Conclusions}

In this paper we have considered the AdS/CFT conjecture for finite N
conformal theories. As opposed to the large N limit which relates the
corresponding
field theories and the classical supergravity, for finite N we need to
consider quantum gravity. We have taken the point of view that the finite
N effects can be captured by gravity on a q-deformed version of 
$AdS_{5} \times S^{5}$ space time where q is a root of unity along the
lines of \ncg.
  We have
discussed mainly the deformation of 
$S^5$ into $S^5_{q}$ where q is a root of unity. The KK reduction on this
quantum sphere is truncated at $N$ if 
$q = e^{i \pi \over {N+4}}$. This agrees with the cutoff on the chiral
primaries of finite N conformal field theory obtained on the boundary of
AdS$_5$.

We then discussed  the $\N = 2$ quotient theories and the
corresponding cutoff in the supergravity and field theory.
By considering only the untwisted sectors the result is that the chiral
primaries have the same cutoff as in the initial $\N = 4$
theory. This result could be explained by considering a quotient of the
quantum sphere. We also derived results in field theories regarding the
cutoffs when twisted sectors are taken into account. We began a
discussion of the corresponding quantum space picture.

It would be interesting to extend the results of this paper for other 
orbifolds of $S^5$, for six dimensional field theories obtained on the
boundary of  $AdS_{7}$ or for three dimensional theories  obtained on the    
boundary of  $AdS_{4}$. 
Some evidence for non-commutative gravity in $AdS_{7} \times S^{4}$
background has been obtained in \bv\ and connections to the quantum group
approach will be interesting to explore.

\noindent{ \bf Acknowledgements : } 
 We are happy to acknowledge fruitful 
 discussions with Antal Jevicki, Gilad Lifschytz,
 and Vipul Periwal. P.M.H. was supported in part by 
 the National Science Council, Taiwan, R.O.C., and
 he thanks C.-S. Chu for discussions and the hospitality of
 the University of Neuch\^{a}tel, Switzerland,
 where part of this work was done.
 The work of R.T. and S.R. was supported
 by   DOE grant  DE-FG02/19ER40688-(Task A). 
 S.R would like to thank the  National Taiwan University 
 for hospitality while part of this work was done.

\listrefs
\end